\documentclass[useAMS,usenatbib]{mn2e}

\usepackage{graphicx}\usepackage{epsfig}\usepackage{epsf}
\usepackage{amsmath}\usepackage{amssymb}\usepackage{stfloats}
\title[UGC~2773-OT: A True $\eta$ Car Analog]{The Persistent Eruption
  of UGC~2773-OT: Finally, a Decade-Long Extragalactic Eta Carinae
  Analog}

\author[Smith et al.]{Nathan Smith$^{1}$\thanks{E-mail:
    nathans@as.arizona.edu}, Jennifer E.\ Andrews$^1$, Jon C.\
  Mauerhan$^{2}$, WeiKang Zheng$^2$, \newauthor
  Alexei V.\ Filippenko$^{2}$, Melissa L. Graham$^2$, and Peter Milne$^1$
  \\ $^{1}$Steward Observatory, University
  of Arizona, 933 N. Cherry Ave., Tucson, AZ 85721, USA \\
  $^2$Department of Astronomy, University of California, Berkeley, CA
  94720-3411, USA}

\begin{document}

\pagerange{\pageref{firstpage}--\pageref{lastpage}} \pubyear{2012}
\maketitle
\label{firstpage}

\begin{abstract}

  While supernova (SN) impostors resemble the Great Eruption of $\eta$
  Carinae in the sense that their spectra show narrow H lines and they
  have typical peak absolute magnitudes of $-$13 to $-$14 mag, most
  extragalactic events observed so far are quite different from
  $\eta$~Car in duration.  Their bright phases typically last for
  $\sim$100~d or less, rather than persisting for several years.  The
  transient object UGC~2773-OT (discovered in 2009) had a similar peak
  absolute magnitude to other SN impostors, but with a gradual 5-yr
  prediscovery rise.  In the $\sim$6 yr since discovery, it has faded
  very slowly (0.26 mag yr$^{-1}$).  Overall, we suggest that its
  decade-long eruption is so far the best known analog of $\eta$~Car's
  19th century eruption.  We discuss extensive spectroscopy of the
  ongoing eruption.  The spectra show interesting changes in velocity
  and line shape that we discuss in detail, including an asymmetric
  H$\alpha$ emission line that we show is consistent with the ejection
  of a bipolar nebula that could be very much like the Homunculus of
  $\eta$ Car.  Moreover, changes in the line width, line profile, blue
  excess emission resembling that of Type~IIn supernovae, and the
  intensity of H$\alpha$ suggest the presence of strong circumstellar
  interaction in the eruption at late times.  This supports the
  hypothesis that the extended plateau of $\eta$ Car's eruption may
  have been powered by shock interaction as well.  One interesting
  difference compared to $\eta$ Car, however, is that UGC~2773-OT so
  far does not exhibit the repeated brief spikes in luminosity that
  have been associated with binary periastron events.

\end{abstract}


\begin{figure*}
\includegraphics[width=6.3in]{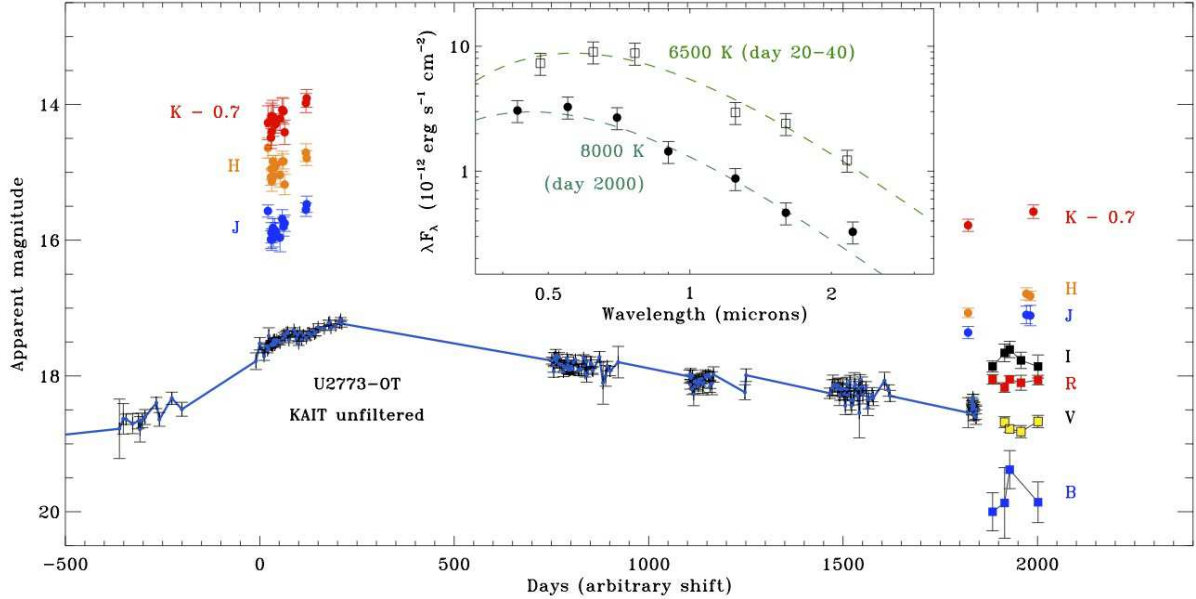}
\caption{The observed apparent magnitude light curve of UGC~2773-OT in
  multiple bands, including the KAIT unfiltered light curve
  (calibrated as $R$), the early $JHK$ points from \citet{smith10},
  late-time $BVRI$ from the Kuiper telescope, and $JHK$ measurements
  from UKIRT (see Tables~\ref{tab:kait} and \ref{tab:phottab}).  The
  inset shows the dereddened SED at early times and late times.}
\label{fig:phot1}
\end{figure*}

\begin{table}\begin{center}\begin{minipage}{1.6in}
      \caption{Unfiltered KAIT Photometry of UGC~2773-OT}
\scriptsize
\begin{tabular}{@{}lcc}\hline\hline
MJD     &mag    &1$\sigma$ (mag) \\ \hline
%
55051.5	&17.78	&0.12 \\
55061.5	&17.52	&0.09 \\
55812.5	&17.77	&0.05 \\
55816.5	&17.84	&0.10 \\
55817.4	&17.93	&0.09 \\
55818.4	&17.76	&0.10 \\
55820.4	&17.72	&0.06 \\
55822.5	&17.70	&0.09 \\
55824.5	&17.79	&0.05 \\
55826.5	&17.74	&0.06 \\
55830.5	&17.72	&0.06 \\
55832.5	&17.83	&0.06 \\
55836.5	&17.83	&0.05 \\
55842.4	&17.94	&0.07 \\
55843.4	&17.85	&0.07 \\
55844.4	&17.88	&0.07 \\
55851.5	&17.74	&0.06 \\
55852.5	&17.89	&0.08 \\
55853.5	&17.84	&0.06 \\
55854.4	&17.78	&0.06 \\
55855.5	&17.90	&0.06 \\
55856.5	&17.91	&0.06 \\
55857.5	&17.90	&0.08 \\
55858.5	&17.89	&0.06 \\
55859.5	&17.89	&0.06 \\
55860.5	&17.91	&0.06 \\
55862.4	&17.86	&0.07 \\
55864.5	&17.88	&0.06 \\
55866.5	&17.90	&0.06 \\
55868.4	&17.80	&0.03 \\
55873.3	&17.76	&0.06 \\
55880.4	&17.92	&0.05 \\
55882.4	&17.91	&0.08 \\
55882.4	&17.91	&0.08 \\
55888.4	&17.90	&0.09 \\
55894.4	&17.73	&0.05 \\
55898.3	&17.96	&0.06 \\
55901.3	&17.79	&0.08 \\
55902.2	&17.99	&0.10 \\
55912.3	&17.81	&0.09 \\
55914.3	&17.83	&0.06 \\
55916.2	&17.91	&0.05 \\
55918.3	&17.92	&0.05 \\
55935.2	&17.72	&0.08 \\
55944.2	&18.10	&0.31 \\
55951.1	&18.00	&0.15 \\
55955.1	&17.85	&0.07 \\
55962.2	&17.93	&0.07 \\
55983.2	&17.80	&0.23 \\
56166.5	&18.01	&0.09 \\
56168.5	&18.13	&0.07 \\
56169.5	&17.97	&0.08 \\
56170.5	&18.03	&0.06 \\
56173.5	&18.19	&0.08 \\
56174.5	&18.00	&0.08 \\
56177.5	&18.27	&0.17 \\
56178.5	&18.09	&0.09 \\
56179.5	&18.16	&0.11 \\
56182.5	&18.07	&0.07 \\
56184.5	&18.10	&0.09 \\
56189.5	&18.14	&0.08 \\
56190.5	&18.08	&0.05 \\
56191.4	&18.05	&0.07 \\
56194.5	&18.09	&0.06 \\
56198.4	&18.04	&0.08 \\
56200.5	&18.12	&0.13 \\
56201.5	&18.12	&0.16 \\
56203.4	&18.09	&0.08 \\
56207.4	&17.99	&0.08 \\
56208.5	&18.02	&0.08 \\
56214.5	&17.98	&0.04 \\
56215.4	&18.00	&0.07 \\
56218.5	&18.01	&0.07 \\
56219.4	&18.16	&0.09 \\
56220.4	&18.12	&0.06 \\
56224.5	&18.11	&0.17 \\
56228.3	&17.98	&0.11 \\
\hline
\end{tabular}\label{tab:kait}\end{minipage}
\end{center}
\end{table}

\begin{table}\begin{center}\begin{minipage}{1.8in}
      \contcaption{Unfiltered KAIT Photometry of UGC~2773-OT}
\scriptsize
\begin{tabular}{@{}lcc}\hline\hline
MJD     &mag    &1$\sigma$ (mag) \\ \hline
56309.1	&18.24	&0.11 \\
56312.1	&17.99	&0.09 \\
56527.5	&18.25	&0.08 \\
56533.5	&18.18	&0.12 \\
56535.5	&18.13	&0.15 \\
56540.5	&18.15	&0.09 \\
56547.5	&18.19	&0.10 \\
56548.5	&18.13	&0.10 \\
56549.5	&18.15	&0.06 \\
56552.4	&18.28	&0.16 \\
56554.4	&18.19	&0.16 \\
56558.4	&18.20	&0.09 \\
56561.4	&18.19	&0.13 \\
56563.5	&18.27	&0.09 \\
56567.4	&18.44	&0.13 \\
56570.5	&18.20	&0.07 \\
56571.4	&18.12	&0.11 \\
56573.4	&18.28	&0.07 \\
56579.4	&18.08	&0.09 \\
56581.4	&18.36	&0.14 \\
56583.4	&18.44	&0.13 \\
56588.4	&18.36	&0.07 \\
56590.5	&18.16	&0.21 \\
56592.4	&18.12	&0.07 \\
56598.4	&18.29	&0.07 \\
56601.4	&18.18	&0.10 \\
56603.4	&18.54	&0.37 \\
56605.5	&18.06	&0.08 \\
56610.2	&18.16	&0.14 \\
56612.3	&18.27	&0.14 \\
56620.4	&18.17	&0.09 \\
56625.3	&18.47	&0.11 \\
56627.3	&18.35	&0.13 \\
56636.3	&18.27	&0.10 \\
56639.2	&18.33	&0.11 \\
56668.2	&18.07	&0.13 \\
56679.2	&18.21	&0.08 \\
56682.2	&18.29	&0.08 \\
56883.5	&18.54	&0.22 \\
56893.4	&18.42	&0.06 \\
56893.5	&18.49	&0.08 \\
56895.5	&18.40	&0.05 \\
56896.5	&18.47	&0.05 \\
56898.5	&18.44	&0.05 \\
56903.5	&18.54	&0.10 \\
\hline
\end{tabular}\end{minipage}
\end{center}
\end{table}

\section{INTRODUCTION}

Observations indicate that there is a population of eruptive or
explosive visible-wavelength transient sources that are systematically
less luminous than supernovae (SNe), but more luminous than any stable
star.  Historically \citep{humphreys99}, four examples were known:
P~Cygni in 1600 AD, $\eta$ Carinae's 19th century eruption (both in
the Milky Way), SN~1954J in NGC~2403 (also known as Variable 12), and
SN~1961V in NGC~1058.  These were considered to be giant eruptions of
luminous blue variable stars (LBVs).  SN~1961V has since
\citep{smith+11,kochanek+11} turned out to be a likely core-collapse
SN of Type~IIn (SN~IIn), but among the others, $\eta$ Carinae's
eruption is often regarded as the prototype of these giant LBV
eruptions, albeit the most extreme example.

Modern surveys for SNe have serendipitously found a few dozen of these
luminous eruptive stars that were sometimes (especially in earlier
studies) considered to be analogs of $\eta$ Carinae and the LBVs.
Because they are typically found in systematic surveys intended to
find SNe, they are often called ``SN impostors'' \citep{vdm12},
although they have also been referred to as $\eta$ Car analogs, Type V
SNe, intermediate-luminosity transients, and other names
\citep{goodrich89,vandyk00,vandyk05,zwicky61,zwicky65}.  See
\citet{smith+11} and \citet{vdm12} for recent general reviews of these
eruptive sources, as well as many references therein regarding
individual objects.  \citet{kochanek11} has discussed considerations
of the important role of dust in these events.

As more examples of these SN impostors have been discovered, and as
more examples have been studied in detail, it has become increasingly
clear that there is a wide diversity among SN impostors, and that few
of them are actually close analogs of $\eta$ Car's historical
eruption.  Some classic SN impostors have luminous, hot, blue
progenitors and an indication that the star survived the eruption,
such as SN~1997bs \citep{vandyk00} and SN~2009ip\footnote{SN~2009ip
  later appeared to explode as a SN in 2012 \citep{mauerhan13}, but
  the star survived its pre-2012 eruptions.}  \citep{smith10,foley11}.
A somewhat different subset of SN impostors includes SN~2008S, the
2008 optical transient in NGC~300, SN~2010dn, and similar objects that
are sometimes referred to as intermediate-luminosity red transients
because of the cooler photospheres that develop
\citep{berger09,bond09,kochanek+11,kochanek+12,thompson09,prieto+08,
  smith+09,smith+11,wesson10}.  Most of these exhibit characteristic
bright emission in the infrared (IR) lines of Ca~{\sc ii} and [Ca~{\sc
  ii}], perhaps related to their dusty circumstellar matter (CSM)
\citep{prieto+08,smith+11}. The most interesting property of these is
that there are a few cases where a deeply dust-enshrouded and
relatively low-luminosity progenitor (perhaps a super-asymptotic giant
branch [AGB] star) has been identified in pre-eruption IR data
\citep{prieto08,prieto+08,thompson09}.  Some objects share properties
of both categories, like UGC~2773-OT (the object discussed in this
paper), which had a luminous blue progenitor and developed the
[Ca~{\sc ii}] and Ca~{\sc ii} emission that is characteristic of the
SN~2008S-like transients \citep{smith10,smith+11,foley11}.  SN~2002bu
shows properties of both classes as well, changing as it faded
\citep{smith+11,skd12}.  SN Hunt248 was an LBV-like eruption that
appeared to arise from a yellow hypergiant progenitor
\citep{mauerhan15}, and SN~2003gm may have been as well
\citep{maund06}.

Additionally, a subset of the SN impostors show very brief (lasting a
few weeks) flares in optical luminosity that fade quickly and in many
cases repeat, and sometimes precede a larger outburst.  Some examples
of these flares are seen in SN~2000ch, SN~1954J before its 1954 peak,
and the pre-SN eruptions of SN~2009ip
\citep{humphreys99,smith10,pastorello10,smith+11,mauerhan13,wagner+04}. Similar
brief peaks were seen in $\eta$ Car's Great Eruption
too\footnote{These ``flashes and relapses'' are what led John Herschel
  to describe $\eta$~Car as ``fitfully variable to an astonishing
  extent'' \citet{herschel1847}.}, and are hypothesised to be related
to stellar collisions at periastron in an eccentric binary system
\citep{smith11}.  Finally, some SN impostors and giant LBV eruptions
have substantially less luminous peaks than $\eta$ Car, with peak
luminosities of $-$10 to $-$11 mag, rather than $-$14 mag.  Examples
of these fainter eruptions are P~Cygni's 17th century eruptions,
SN~1954J, SN~2002kg (although we caution that SN~2002kg may have
actually been a normal S~Dor eruption and not a giant eruption;
\citealt{maund06,vandyk06}), V1 in NGC~2366 \citep{drissen01} and the
1990s eruption of HD~5980 in the Small Magellanic Cloud (see
\citealt{gloria04}).  Thus, there is a very wide diversity among this
class of eruptions, if indeed they belong in the same class.

\citet{kochanek+12} have pointed out that most SN impostors do not
match traditional expectations for LBV giant eruptions as summarised
by \citet{HD94}, and suggest that many of them are more likely to be
SN~2008S-like events.  This may partly be an expression of the fact
that most aspects of the traditional view of LBVs (HD94) have not
stood the test of time.  Continued study has shown that rigid
imposition of definitions of LBVs (HD94) have not been supported by
detailed analysis on several fronts, and these definitions would in
fact exclude most LBVs from the class.  In particular, the important
conjecture that S Dor variations are caused by optically thick winds
at constant $L_{\rm Bol}$ may be wrong, and steady super-Eddington
winds may not be the correct or dominant driving mechanism in many
giant LBV eruptions (see review by \citealt{smith14}).

Studies with quantitative spectroscopy
\citep{dekoter96,groh09a,groh09b} have disproven the conjecture that
S~Dor brightening events are caused by developing pseudo-photospheres
in optically thick winds (\citealt{davidson87}, HD94).  The mass-loss
rates of S~Dor maxima are not high enough to make such large
pseudo-photospheres, and so they are more likely to be caused by
envelope inflation or pulsation \citep{graefner12}.  Moreover,
bolometric luminosities during S~Dor eruptions are not really constant
\citep{groh09a}.  Similarly, the idea that giant-eruption maxima are
caused by pseudo-photospheres in opaque super-Eddington winds is
challenged by light-echo spectra of $\eta$ Carinae
\citep{rest12,prieto14}, by detailed analysis of the ejecta around
$\eta$ Car that are better matched by explosive models
\citep{smith03,smith06,smith08}, and the fact that many extragalactic
giant LBV eruptions are relatively hot at peak luminosity rather than
cool \citep{smith+11,mauerhan15}.  Last, as discussed by \citet{st15},
the central paradigm of the role that LBVs play in stellar evolution
and their connection to stars with the highest initial masses is
probably incorrect, because their isolation from massive O-type stars
dictates that they are largely products of binary evolution and not a
transitional state in the lives of the most massive stars.  There are
many ways in which violent binary interaction may be important for
understanding LBVs and related transients
\citep{soker01,pod10,kochanek+14,st15,smith14}.

\begin{table*}\begin{center}\begin{minipage}{6.4in}
      \caption{Visible and IR Photometry}
\scriptsize
\begin{tabular}{@{}llcccccccccccccc}\hline\hline
%
Date	 &JD		&$B$ &1$\sigma$	&$V$ &1$\sigma$	&$R$ &1$\sigma$	&$I$ &1$\sigma$	&$J$ &1$\sigma$	&$H$ &1$\sigma$	&$K$ &1$\sigma$ \\ \hline 
8/19/14	 &2456883	&...	&...	&...	&...	&...	&...	&...	&...	&17.36	&0.07	&17.07	&0.07	&16.48	&0.09	\\
10/16/14 &2456946	&20.00	&0.14	&...	&...	&18.05	&0.07	&17.86	&0.08	&...	&...	&...	&...	&...	&...	\\
11/16/14 &2456977	&19.87	&0.26	&18.68	&0.08	&18.17	&0.07	&17.66	&0.13	&...	&...	&...	&...	&...	&...	\\
11/29/14 &2456990	&19.38	&0.14	&18.78	&0.05	&18.05	&0.05	&17.61	&0.12	&...	&...	&...	&...	&...	&...	\\
12/28/14 &2457019	&...	&...	&18.82	&0.09	&18.10	&0.11	&17.77	&0.12	&...	&...	&...	&...	&...	&...	\\
1/11/15	 &2457033	&...	&...	&...	&...	&...	&...	&...	&...	&17.10	&0.09	&16.79	&0.10	&...	&...	\\
1/21/15  &2457043	&...	&...	&...	&...	&...	&...	&...	&...	&17.11	&0.07	&16.82	&0.09	&...	&...	\\
1/29/15	 &2457051	&...	&...	&...	&...	&...	&...	&...	&...	&...	&...	&...	&...	&16.28	&0.1	\\
2/10/15	 &2457063	&19.86	&0.15	&18.67	&0.09	&18.06	&0.07	&17.86	&0.17	&...	&...	&...	&...	&...	&...	\\
\hline
\end{tabular}\label{tab:phottab}\end{minipage}
\end{center}
\end{table*}


\begin{figure*}
\includegraphics[width=6.3in]{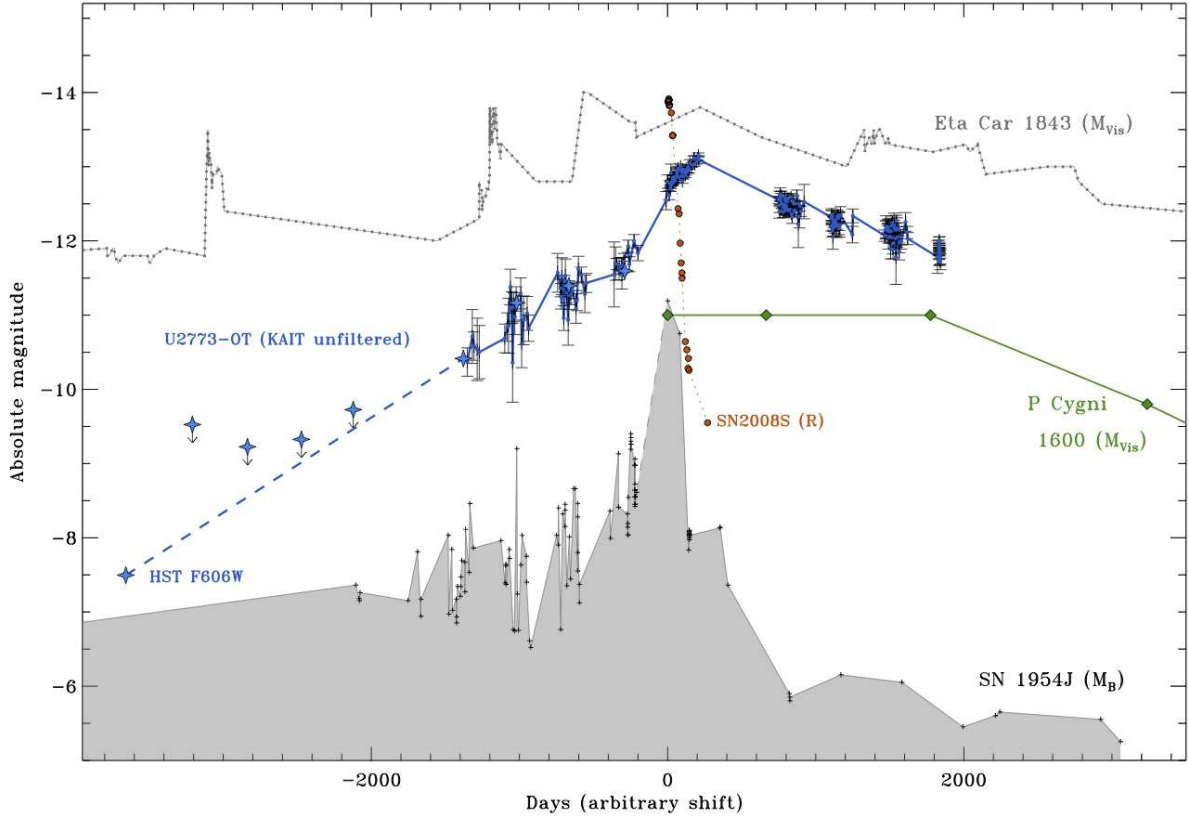}
\caption{The absolute magnitude light curve of UGC~2773-OT compared to
  those of other transients: $\eta$ Car's 19th century eruption
  \citep{sf11}, P Cyg's 1600 AD eruption (Smith et al.\ 2015, in
  prep.), and the brief SN impostor SN~2008S \citep{smith+09}.}
\label{fig:phot}
\end{figure*}

In any case, it is perhaps sufficient to state that our understanding
of LBVs, LBV giant eruptions, shells around nearby LBVs, and
extragalactic SN impostors is still too poor to make broad
generalizations about the nature of the SN impostors or their
connection to LBVs.  The number of well-studied objects is still very
small, and a close look at the physical properties of additional
examples is quite valuable.  Much of our interpretation of these
events is coloured by the vast amount and quality of data for
$\eta$~Car that far surpasses all other cases, making comparisons
tempting but sketchy.  Finding extragalactic analogs that actually
match $\eta$~Car has so far been difficult.

In this paper we present an update on the ongoing outburst of the
LBV-like transient UGC~2773-OT.  The ``update'' refers to developments
since our first paper on this object \citep{smith10}, which discussed
photometry and spectra obtained shortly after the time of discovery,
as well as archival data that showed a relatively faint progenitor 10
years before discovery and a gradual brightening for 5 yr before (see
below).  In that paper, we found that the transient had a relatively
cool spectrum ($\sim$7000~K) and relatively slow outflow speeds
(200--300 km s$^{-1}$), a mild IR excess from circumstellar dust, and
a luminous yellow progenitor with log($L$/L$_{\odot}$) = 5.1 (although
it could have been substantially hotter and more luminous, due to
possible foreground extinction from circumstellar dust).  The likely
effective initial mass indicated by comparing the quiescent progenitor
to single-star models was found to be $\gtrsim$20 M$_{\odot}$ (more if
additional extinction was significant or if the star was hot).
\citet{foley11} presented additional complementary spectra of
UGC~2773-OT, and came to similar conclusions about its physical
properties and LBV-like nature.  In the time since these first papers,
UGC~2773-OT has continued to evolve slowly, declining from its peak
but remaining at relatively high outburst luminosity.  If the
progenitor really was an initially 20 M$_{\odot}$ star, then the
transient has exceeded the classical Eddington luminosity for more
than a decade.  The fact that it has persisted for a decade in this
high-luminosity state is physically meaningful, as discussed in this
paper.  In this sense, we suggest that it is so far the only known
example of an extragalactic SN impostor that behaves similar to the
19th century Great Eruption of $\eta$~Carinae.  This extends the
comparison made by \citet{rest12}, who remarked that spectra of
UGC~2773-OT obtained shortly after discovery provided the closest
match among known SN impostors to the low-temperature spectra of
$\eta$ Car's light echoes.  In this paper we report changes in the
spectrum of UGC~2773-OT after several years, and it will be
interesting to see if $\eta$ Car's light echoes develop similar
changes as time passes.

UGC~2773-OT was observed to occur in the central parts of the dwarf
irregular galaxy UGC~2773, and its brightening was discovered
\citep{boles09} on 2009 August 18.08 (UT dates are used throughout
this paper).  As in our previous paper \citep{smith10}, we adopt $m-M$
= 28.82 mag, $E(B-V)$ = 0.56 mag, and $A_R$ = 1.51 mag for the Milky
Way reddening and extinction in the line of sight to the host galaxy
UGC~2773, and we refer to day 0 as the date of discovery.  We present
new photometry and spectra in Section 2, and we combine these with our
previously published photometry and spectra.  In Section 3 we
summarise the observational results, and in Section 4 we discuss
physical interpretations in context with other LBVs --- especially one
of them.

\section{OBSERVATIONS}

\subsection{KAIT Photometry}

UGC~2773, the dwarf irregular host galaxy of UGC~2773-OT, has been
monitored regularly with the Katzman Automatic Imaging Telescope
(KAIT; \citealt{filippenko03}) at Lick Observatory.  As demonstrated
by \citet{li02}, the best match to broadband filters for the KAIT
unfiltered data is the $R$ band (although note that bright H$\alpha$
line emission will influence the $R$ band more than the unfiltered
data).  We list the KAIT apparent unfiltered magnitudes of UGC~2773-OT
in Table~\ref{tab:kait}, and the apparent magnitudes are plotted in
Figure~\ref{fig:phot1}.  This table lists only KAIT observations since
the time of discovery.  We analyzed prediscovery unfiltered images and
detected a source at the position of UGC~2773-OT during the $\sim$5 yr
before discovery, as well as upper limits before that.  There were
multiple observations each year, so we produced stacked seasonal
averages to improve the sensitivity.  Further information about the
prediscovery photometry is available in our previous paper
\citep{smith10}.  A stacked image from the year 2000, when UGC~2773-OT
was not detected, is used as a template image in an image-subtraction
technique to cleanly remove the galaxy contamination at the position
of UGC~2773-OT in later images.  This is important because UGC~2773-OT
is in a crowded region of its host, and so we may expect different
results from template-subtracted photometry and raw aperture
photometry.  The photometry listed here is a reduction and analysis of
the same data from 2009 that were published earlier, plus new
observations since 2009.  Unfortunately, KAIT observations of
UGC~2773-OT during the 2010 season are not available because of a
failure in the data storage device, so there is a gap in the light
curve 1 yr after discovery.

To put the flux on an absolute magnitude scale, we adopt the same
distance and reddening from our previous study \citep{smith10}, and
listed here in the Introduction.  The resulting absolute magnitude
light curve is shown in Figure~\ref{fig:phot}, along with some other
transient sources from the literature for comparison.

\subsection{Visible/IR Photometry}

Johnson $B$, $V$, Harris $R$, and Arizona $I$ band imaging was
obtained with the Mont4k instrument mounted on the 1.55m Kuiper
telescope located on Mt.\ Bigelow, Arizona. Observations were taken on
2014 October 16, November 16, 29, and December 28, and on 2015
February 10 in 3$\times$3 binning mode, resulting in a final pixel
scale of 0$\farcs$43.  The observations were then reduced and combined
into final images using standard {\sc IRAF} {\tt ccdred} procedures
and dome flats.

NIR imaging observations were taken with the 3.8-m United Kingdom
Infrared Telescope (UKIRT) on Mauna Kea using UFTI (2014 August 19)
and WFCAM2 (2015 January 11, 21, and 29).  $JHK$ observations were
pipeline reduced by the Cambridge Astronomical Survey Unit (CASU).

Aperture photometry was performed on all images using a 1$\farcs$3
aperture to mitigate contamination from adjacent host galaxy light,
and was calibrated using APASS optical and 2MASS NIR standard stars
present in the field.  $R$ and $I$ band standards were transformed
from APASS $r\prime$ and $i\prime$ magnitudes using Jester et al.\
(2005).  For both optical and NIR data, uncertainties were calculated
by adding in quadrature photon statistics and zero point deviation of
the standard stars for each epoch.  The resulting photometry is listed
in Table~\ref{tab:phottab}.

\begin{table}\begin{center}\begin{minipage}{4.0in}
      \caption{Log of Spectroscopic Observations}
\scriptsize
\begin{tabular}{@{}lclcl}\hline\hline
UT date &day$^a$   &Tel./Instr. &$\Delta\lambda$(\AA) &Comment \\ \hline
2009 Sep 22 &34    &Keck/LRIS   &3600-9200            &Paper I, low res \\
2009 Sep 22 &34    &Keck/LRIS   &3800-5100            &Paper I, hi res \\
2009 Sep 22 &34    &Keck/LRIS   &6250-7850            &Paper I, hi res \\
2011 Jan 13 &478   &MMT/BCh     &6345-7640            &1200 lpm \\
2011 Jan 14 &479   &MMT/BCh     &6345-7640            &1200 lpm \\
2011 Jun 27 &492   &MMT/BCh     &6200-7490            &1200 lpm \\
2012 Jan 03 &832   &MMT/BCh     &6350-7645            &1200 lpm \\
2012 Jan 20 &849   &MMT/BCh     &3850-9050            &300 lpm \\
2012 Jan 31 &860   &MMT/BCh     &3850-9050            &300 lpm \\
2012 Aug 14 &1057  &Bok/BC      &3610-6900            &low res \\
2012 Oct 16 &1116  &MMT/BCh     &5700-7000            &1200 lpm \\
2012 Nov 24 &1124  &MMT/BCh     &5700-7000            &1200 lpm \\
2012 Nov 25 &1125  &MMT/BCh     &3850-9050            &300 lpm \\
2013 Sep 04 &1443  &MMT/BCh     &5700-7000            &1200 lpm \\
2013 Dec 29 &1559  &MMT/BCh     &3850-9050            &300 lpm \\
2014 Jan 03 &1564  &MMT/BCh     &5700-7000            &1200 lpm \\
2014 Apr 08 &1659  &MMT/BCh     &5700-7000            &1200 lpm \\
2014 Oct 02 &1836  &Keck/DEIMOS &4410-9640            &low res \\
2014 Oct 15 &1849  &MMT/BCh     &5700-7000            &1200 lpm \\
2015 Mar 22 &2007  &MMT/BCh     &5700-7000            &1200 lpm \\
\hline
\end{tabular}\label{tab:speclog}\end{minipage}
\end{center}
$^a$Relative to discovery (as in Paper I).
\end{table}

\begin{figure*}
\includegraphics[width=5.2in]{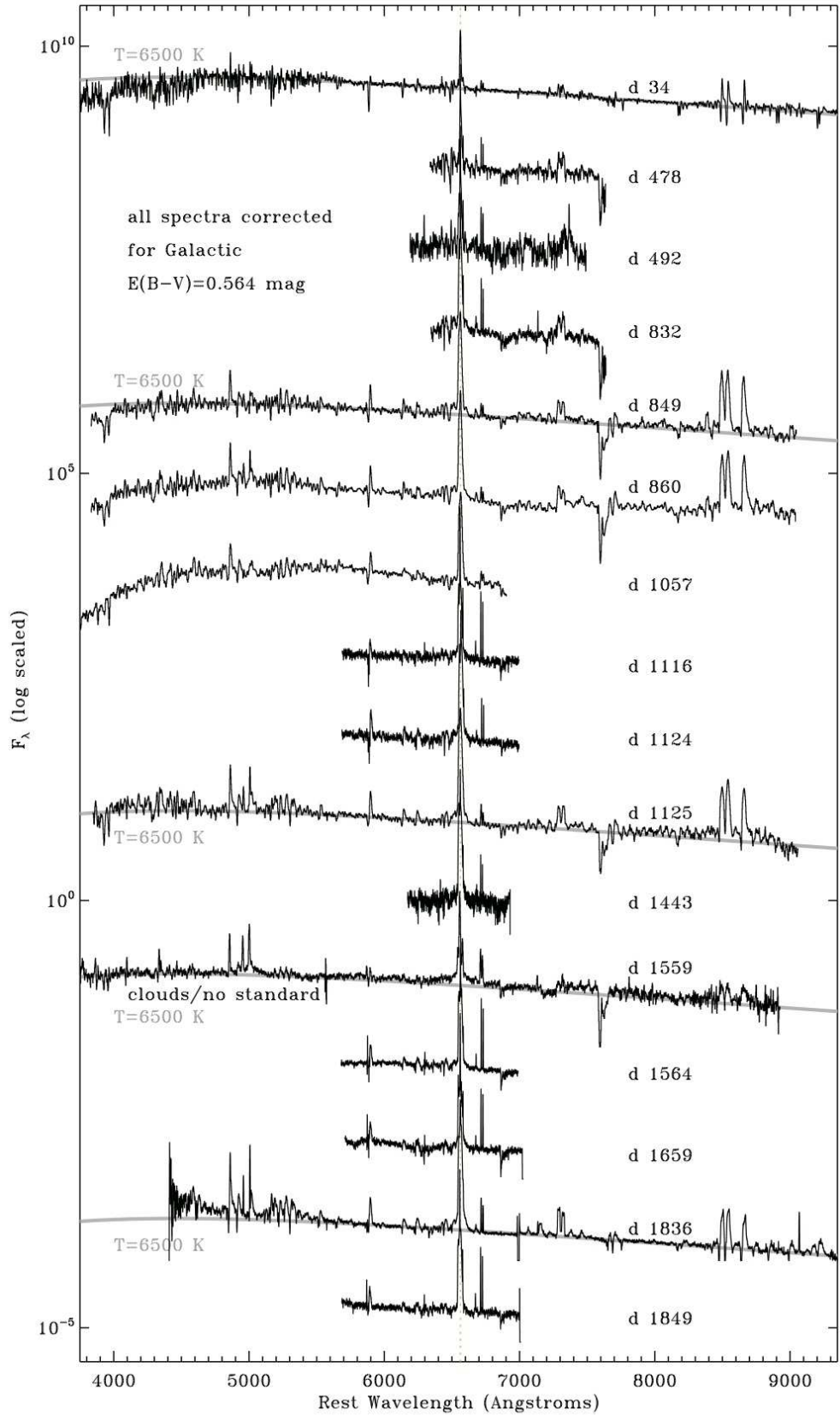}
\caption{Visible-wavelength spectra of UGC~2773-OT obtained with the
  Keck, MMT, and Bok telescopes over the past 6 years (see
  Table~\ref{tab:speclog}).  As noted in the figure, all spectra shown
  here have been corrected for the Milky Way line-of-sight reddening.}
\label{fig:spec}
\end{figure*}

\subsection{Spectra}

\begin{figure}
\includegraphics[width=3.2in]{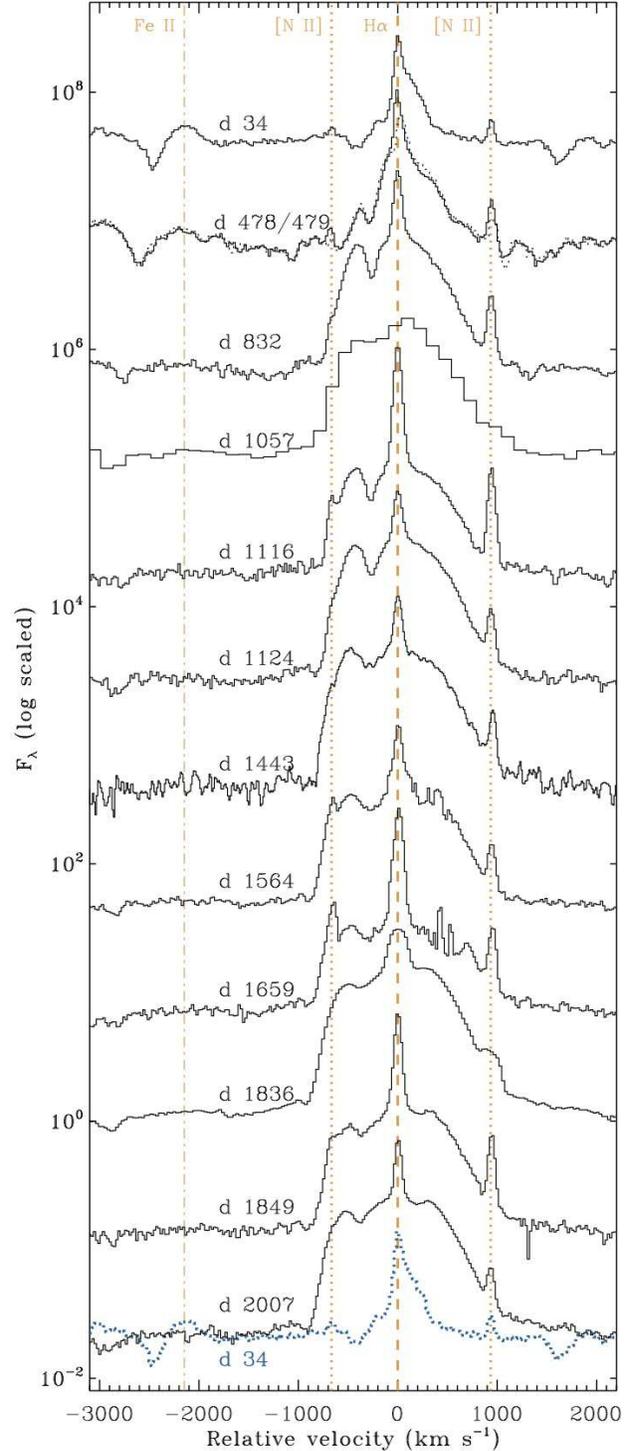}
\caption{The H$\alpha$ line profile as seen in our spectra
  of UGC~2773-OT.  Narrow emission components of [N~{\sc ii}] and
  H$\alpha$ that are probably associated with an underlying H~{\sc ii}
  region or distant CSM are indicated by the orange vertical lines
  through the figure.  An orange vertical line also notes the location
  of an Fe~{\sc ii} line.  The first-epoch spectrum (day 34) is
  reproduced at the bottom of the plot in a dashed-blue tracing for
  comparison with the latest epochs.}
\label{fig:halpha}
\end{figure}

\begin{figure}
\includegraphics[width=2.8in]{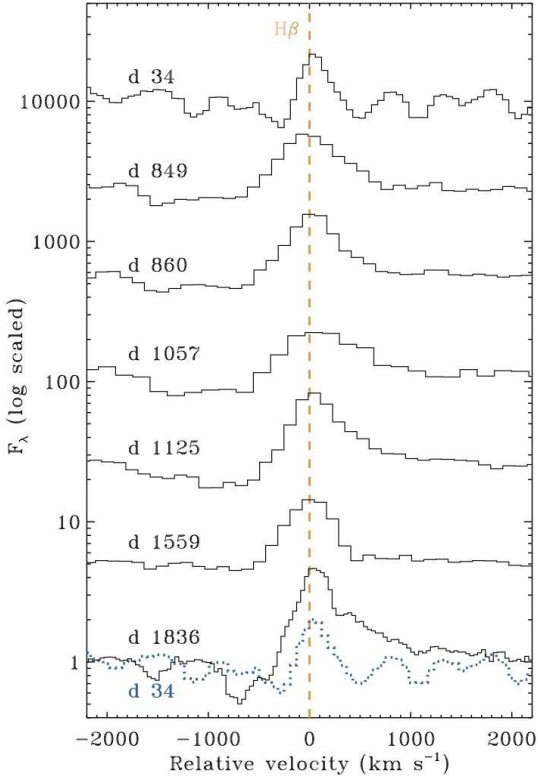}
\caption{Same as Figure~\ref{fig:halpha}, but for H$\beta$.  Our
  spectral dataset includes fewer epochs that sample H$\beta$, and
  only with low-resolution spectra that cover a broader wavelength
  range.}
\label{fig:hbeta}
\end{figure}

\begin{figure}
\includegraphics[width=2.9in]{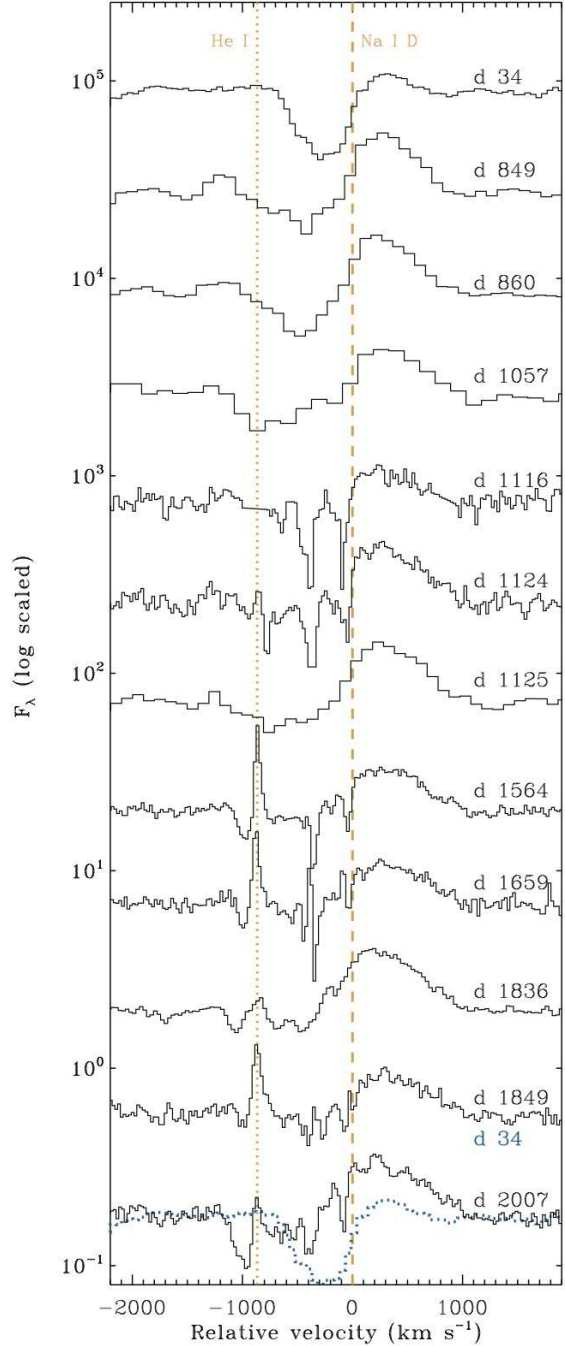}
\caption{Line profiles of Na~{\sc i}~D (at v=0 km
  s$^{-1}$) and He~{\sc i} $\lambda$5876 as seen in our spectra of
  UGC~2773-OT.  The first epoch spectrum (day 34) is reproduced at the
  bottom of the plot in a dashed blue tracing for comparison with the
  latest epochs.}
\label{fig:nad}
\end{figure}

\begin{figure}
\includegraphics[width=2.9in]{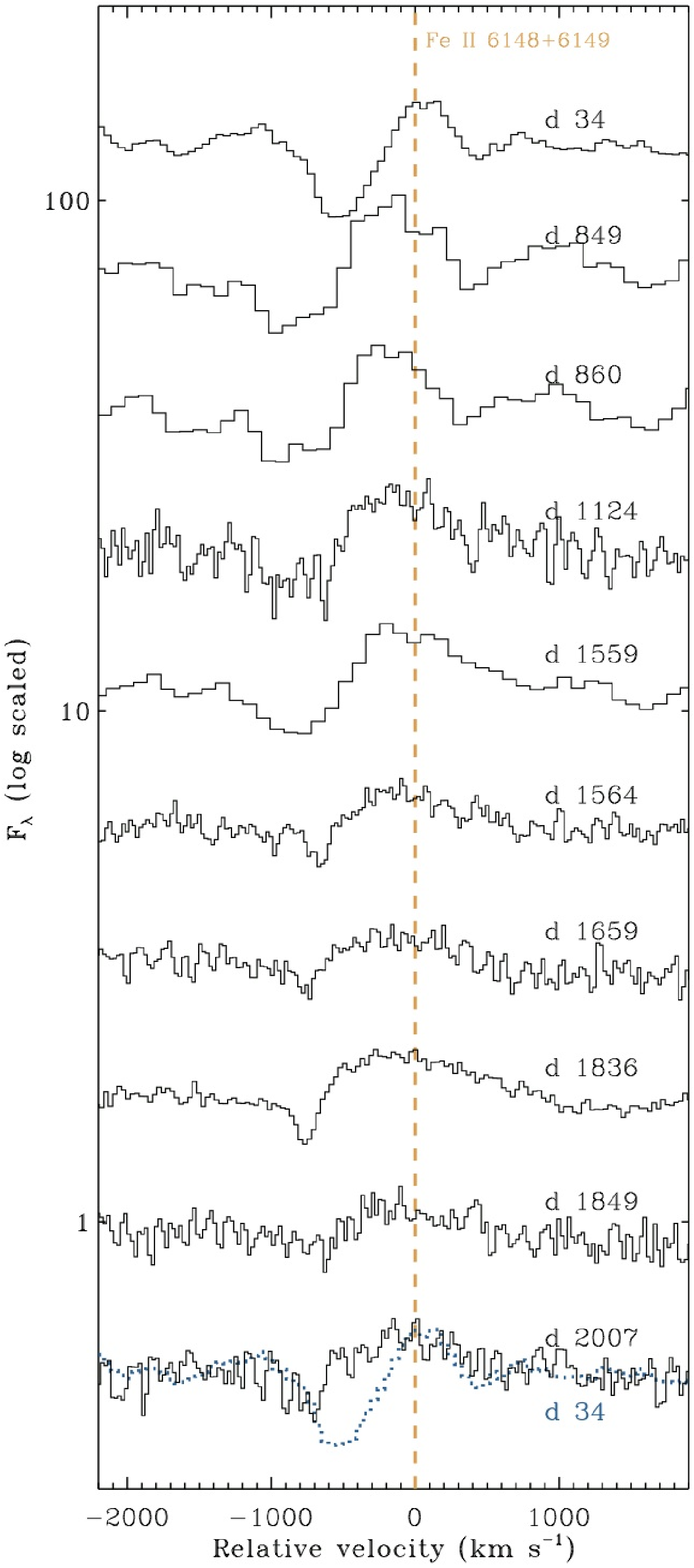}
\caption{The Fe~{\sc ii} $\lambda\lambda$6148,6149 line profile as
  seen in our spectra of UGC~2773-OT.  The first epoch spectrum (day
  34) is reproduced at the bottom of the plot in a dashed blue tracing
  for comparison with the latest epochs.}
\label{fig:fe2}
\end{figure}

We obtained several epochs of optical spectroscopy of UGC~2773-OT
using a number of different facilities over the past 6 years,
including the Bluechannel (BC) spectrograph on the 6.5-m Multiple
Mirror Telescope (MMT), the Boller \& Chivens (B\&C) Spectrograph
mounted on the 2.3-m Bok telescope on Kitt Peak, the Low-Resolution
Imaging Spectrometer (LRIS; \citealt{oke95}) mounted on the 10-m
Keck~I telescope, and the Deep Imaging Multi-Object Spectrograph
(DEIMOS; \citealt{faber03}) on Keck~II.  Details of the spectral
observations are summarised in Table~\ref{tab:speclog}.  For MMT
spectra, we either used the 1200 lpm grating (moderately high
resolution) or the 300 lpm grating (low resolution).  The slit was
always oriented at the parallactic angle \citep{filippenko82}, and the
long-slit spectra were reduced using standard procedures.  Final
spectra are shown in Figure~\ref{fig:spec}, including the first epoch
of spectra, which is from our previous paper \citep{smith10}.  Several
epochs in Figure~\ref{fig:spec} have a blackbody plotted in grey;
these are intended only as a rough relative comparison.  Nevertheless,
inferred temperatures around 6500~K suggest that these quoted
temperatures are not wildly in error.  Details of the H$\alpha$ line
profile are shown in Figure~\ref{fig:halpha}, H$\beta$ is shown in
Figure~\ref{fig:hbeta}, details of the region around Na~{\sc i} D and
He~{\sc i} $\lambda$5876 are shown in Figure~\ref{fig:nad}, and
Fe~{\sc ii} $\lambda\lambda$6148,6149 is shown in
Figure~\ref{fig:fe2}.

\section{RESULTS}

\subsection{Light Curve}

In our previous study \citep{smith10}, we showed that prediscovery
photometry of UGC~2773-OT showed a slow and steady brightening in the
5~yr prior, when it was 3 or more magnitudes brighter than its
quiescent progenitor detected a decade before discovery in {\it HST}
images (the prediscovery brightening is reproduced in
Figure~\ref{fig:phot}).  As noted earlier \citep{smith10,foley11}, the
{\it HST} data for the progenitor correspond to a star of at least
$\sim$20 M$_{\odot}$ and log($L$/L$_{\odot}$) = 5.1.  Those authors both
noted, however, that these estimates for the progenitor are likely to
be just lower limits because with a small amount of circumstellar dust
extinction (motivated by the early IR excess), the progenitor could
easily be hotter and more luminous, corresponding to a 40--60
M$_{\odot}$ star or more.  Thus, the slow brightening in the years
before discovery could have been partially due to colour evolution from
a hot to a cooler state, in addition to a real increase in bolometric
luminosity.  Unfortunately, information about the prediscovery colour
evolution is not available aside from the single {\it HST} epoch.

Since reaching its peak apparent brightness around day 200 (as plotted
in Figure~\ref{fig:phot}) until day $\sim$2000, UGC~2773-OT has had a
very slow and steady decline rate at visible wavelengths.  There may
be small fluctuations in brightness, but the overall trend appears
smooth.  From the KAIT light curve, we measure an average decline rate
of 0.26 mag yr$^{-1}$ during the first $\sim$2000 days.

There is very little colour evolution during this slow decline, and the
trend is toward slightly bluer colours at late times.  In the inset of
Figure~\ref{fig:phot1}, we show the spectral energy distribution (SED)
for early times around days 20--40, as well as late times around day
2000.  The early-time photometry include IR $JHK$ magnitudes from our
previous study \citep{smith10} and $gri$ magnitudes from
\citet{foley11} taken at the same epoch, both dereddened by
$E(B-V)$=0.56 mag.  The optical/IR SED at early times can be
approximated with a $\sim$6500~K blackbody, which is consistent with
our low-resolution spectrum on day 34 shown in Figure~\ref{fig:spec}.
At later times, the SED indicates a slightly warmer temperature, but
as we discussed below, this is mostly due to a combination of a
continuum photosphere at about the same temperature, plus a blue
excess from emission lines.

Neither early nor late epochs show a substantial IR excess.  At early
times, one could deduce a small near-IR excess by choosing a slightly
warmer characteristic temperature for optical data plus a small amount
of hot dust to match the IR fluxes \citep{foley11,smith10}.  At late
epochs, the $J$ and $H$ magnitudes are consistent with an extension of
the optical blackbody, but there does appear to be a small $K$-band
excess.  This may be due to the formation of new dust, cooling of CSM
dust heated by an IR echo, or the destruction of the hottest dust seen
previously, leaving the cooler dust at larger radii.  Continued
monitoring of the IR fluxes may be interesting; if UGC2773-OT fades as
$\eta$~Carinae did at the end of its decade-long plateau, we may see a
strong increase in the IR excess.

Although there has been only gradual change in the light curve of
UGC~2773-OT over the past few years, there have been some pronounced
changes in its spectrum, as discussed next.

\begin{figure}
\includegraphics[width=3.0in]{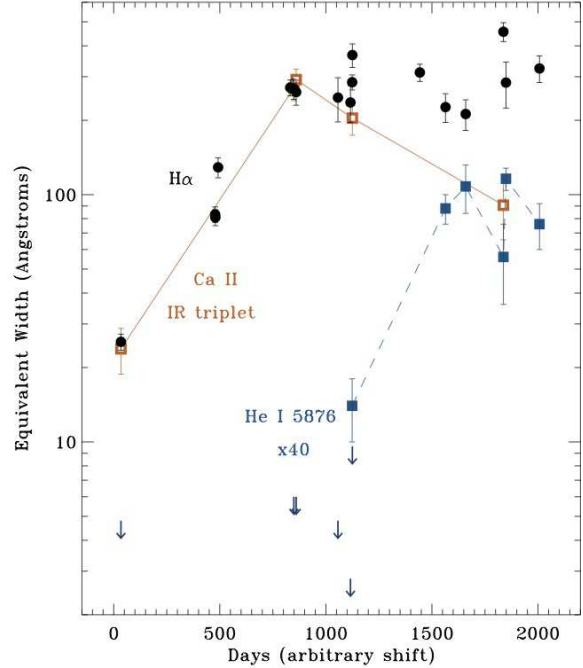}
\caption{Measured emission-line EWs (emission is positive) of
  H$\alpha$ as compared to the EWs for the Ca~{\sc ii} IR triplet and
  He~{\sc i} $\lambda$5876.  H$\alpha$ (black filled points) was
  integrated from roughly $-$1200 to $+$1800 km s$^{-1}$, and the
  uncertainty is dominated by signal to noise of the continuum.  For
  Ca~{\sc ii} (orange unfilled squares, connected by orange lines), we
  took the integrated EW of all 3 lines, and so P Cygni absorption at
  some epochs weakens the emission EW.  The relatively wide continuum
  also raises the EW uncertainty due to the choice of the continuum
  level.  For He~{\sc i} $\lambda$5876 (blue squares and arrows) we
  have only upper limits at early times determined by the signal to
  noise ratio of the spectra, whereas we clearly detect narrow He~{\sc
    i} at later times, although at those epochs we sometimes also see
  an accompanying P Cygni absorption that adds to uncertainty in
  choosing a continuum level. (We did not use the day 1559 spectrum,
  which suffered from calibration problems due to weather.) The
  He~{\sc i} EW is multipled by a factor of 40 for display.}
\label{fig:ew}
\end{figure}

\subsection{Spectroscopy}

In our previously reported day 34 spectrum \citep{smith10} obtained
shortly after discovery and near the time of peak luminosity,
UGC~2773-OT showed a cool absorption-line spectrum, similar to late
F-type of G-type supergiants, and similar in some ways to the spectra
of LBVs in their cool state.  \citet{foley11} presented similar
spectra of UGC~2773-OT obtained around the same epoch. The early
spectrum was characterised by heavy line blanketing by numerous narrow
absorption lines in the blue, and relatively weak and narrow H$\alpha$
emission.  An echelle spectrum taken with Keck/HIRES on day 22 showed
a clear P Cygni profile in H$\alpha$ that suggested outflow speeds
around 360 km s$^{-1}$ \citep{smith10}.  The early spectrum showed
strong absorption from Ca~{\sc ii} H and K, very weak emission in the
forbidden [Ca~{\sc ii}] doublet, and a relatively weak P Cygni profile
in the Ca~{\sc ii} IR triplet \citep{smith10}. These are similar to
the spectral properties seen in SN~2008S-like objects
\citep{prieto08,prieto+08,thompson09}.  Subsequently, \citet{rest12}
showed that our day 34 spectrum of UGC~2773-OT resembled spectra of
light echoes from $\eta$~Carinae, also thought to correspond to early
phases in its historical giant LBV eruption near peak.

In the intervening $\sim$6 yr, the spectrum of UGC~2773-OT has shown
gradual and subtle changes as the source faded slowly.  The overall
spectral evolution is shown in Figure~\ref{fig:spec}. While the
visible/red continuum slope has remained roughly constant at a
characteristic apparent temperature around 6500~K, there are a number
of other changes reminiscent of increasing temperature/ionization or
dropping optical depth.  The forest of narrow absorption lines that
cause strong line blanketing in the blue (day 34) weakened
considerably, and many narrow features including numerous Fe~{\sc ii}
lines change into emission rather than absorption; this is discussed
more below.  We see weakening and disappearing P Cygni absorption
features, and strengthening with time of emission components of Balmer
lines and the Ca~{\sc ii} triplet.  Later epochs even show narrow
He~{\sc i} in emission, which is absent at early epochs.  One of the
most pronounced changes is seen in the H$\alpha$ profile, which
becomes stronger, broader, and develops an asymmetric and irregular
profile (Figure~\ref{fig:halpha}).  Implications of the H$\alpha$
profile are discussed later.\footnote{Since our spectra were obtained
  at epochs that still trace relatively high densities in the outflow,
  forbidden lines that one normally uses to study abundances are
  mostly quashed (and would be difficult to separate from a
  surrounding H~{\sc ii} region anyway), so we do not provide a
  discussion of the CNO composition of UGC~2773-OT for comparison with
  $\eta$ Car \citep{davidson86,smithmorse04}.}

An interesting comparison involves the relative strengths of
H$\alpha$, the Ca~{\sc ii} IR triplet, and He~{\sc i} $\lambda$5876.
The time dependence of equivalent widths (EWs) for these three lines
is plotted in Figure~\ref{fig:ew}.  While the light curve shows only
gradual fading of about 1.5 mag (a factor of 4) in $\sim$2000 days,
the H$\alpha$ EW climbs by a factor of $\sim$150 or more, so these
changes in EW also trace changes in line luminosity.\footnote{In fact,
  a high resolution spectrum on day 22 \citep{smith10} shows that more
  than 2/3 of the early time H$\alpha$ EW we measure is actually due
  to narrow H$\alpha$ and [N~{\sc ii}] from an unresolved H~{\sc ii}
  region, whereas the H~{\sc ii} region contributes a much lower
  fraction at late times because the broad component has strengthened.
  The true relative increase in H$\alpha$ strength is therefore even
  larger than indicated in Figure~\ref{fig:ew} by another factor of
  2$-$3.}

Underscoring the unusual changes to H$\alpha$, the H$\beta$ line does
not show a similar evolution in terms of its profile shape or relative
strength.  On day 34, the H$\alpha$/H$\beta$ flux ratio is 7.7$\pm$1
(not corrected for any local reddening), and the H$\alpha$/H$\beta$
flux ratio then climbs to 31$\pm$2 by day 1057.  This indicates that
H$\beta$ does not increase in strength as much as H$\alpha$, probably
signifying a growing contribution from shock excitation at later
times. This increasing H$\alpha$/H$\beta$ ratio is not due to
increased reddening from dust, since the continuum shows basically no
change in slope, while H$\alpha$ develops a line profile that is very
different from H$\beta$.  Also, the profile of H$\beta$ retains its P
Cyg absorption, whereas H$\alpha$ goes fully into emission with an
asymmetric profile.  Interestingly, though, the speed of the H$\beta$
P Cygni trough increases from around 300 km s$^{-1}$ on day 34 to
about 700--800 km s$^{-1}$ by day 1836 (Figure~\ref{fig:hbeta}).  An
important point is that at each epoch, the blue P Cygni absorption in
H$\beta$ is at a similar speed to the blue emission bump seen in
H$\alpha$.  Thus, both H$\beta$ and H$\alpha$ trace a dramatic
increase in outflow speed with time during the event.  If the outflow
speed is increasing with time, the fast material must catch up with
and overtake the slower material ejected earlier, indicating that
shocks should play an important role in the event.  This is discussed
more below.

The EW of the Ca~{\sc ii} IR triplet increases in-step with H$\alpha$
for the first 1000 days, but while H$\alpha$ levels off for the next
1000 days, Ca~{\sc ii} fades by more than a factor of 3.  At around
the same time, He~{\sc i} $\lambda$5876 transitions from nondetection
in the first 1000 days, to then rising as Ca~{\sc ii} falls (note that
the He~{\sc i} EW is multiplied by a factor of 40 for display in
Figure~\ref{fig:ew}).  Thus, the changes after day 1000 likely
indicate a rise in ionization and electron temperature (especially
He~{\sc i} emission, but also Ca$^+$ getting partly ionized to
Ca$^{++}$) that accompanies a falling optical depth.  The likely
physical significance of these changes is discussed below in Section
4.1.

\begin{figure}
\includegraphics[width=3.0in]{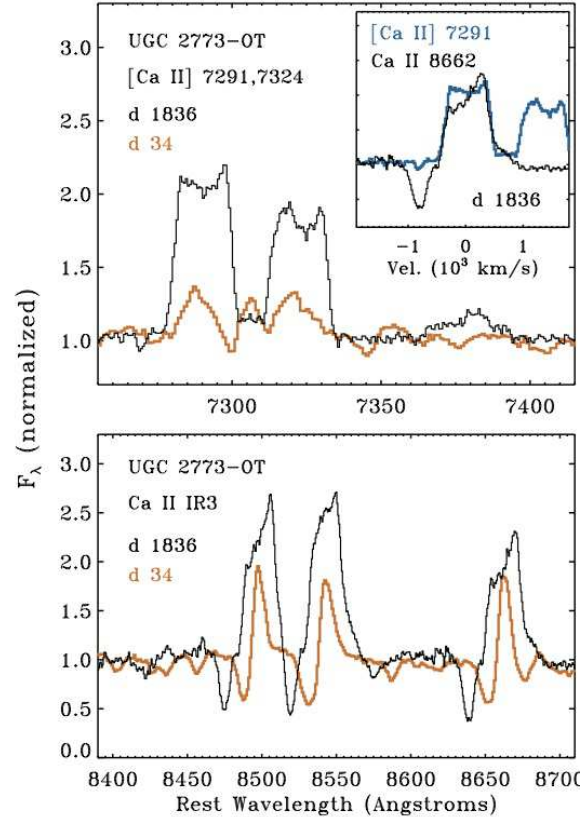}
\caption{A comparison of the early and late-time spectra for
  UGC~2773-OT, highlighting the red Ca$^+$ lines.  The top panel shows
  the [Ca~{\sc ii}] $\lambda\lambda$7291,7324 doublet on days 34
  (black) and 1836 (orange), while the bottom panel shows the same two
  dates for the Ca~{\sc ii} IR triplet at 8498, 8542, and 8662 \AA.
  Both sets of lines show an increase in strength, and the P Cygni
  absorption of the Ca~{\sc ii} IR triplet shows an unambiguous
  increase in velocity. The inset in the upper right shows the line
  profile shapes as a function of rest-frame velocity for [Ca~{\sc
    ii}] $\lambda\lambda$7291 (blue) and Ca~{\sc ii} $\lambda$8662
  (black).}
\label{fig:ca2}
\end{figure}

The behavior of the red/IR lines of Ca$^+$ are particularly
interesting, and are relevant to comparisons with other types of SN
impostors and $\eta$~Car.  Figure~\ref{fig:ca2} illustrates the
changes in line intensity and profile shape with time.  This figure
shows the day 34 (near peak) and day 1836 spectra of the [Ca~{\sc ii}]
$\lambda\lambda$7291,7324 doublet and the Ca~{\sc ii} IR triplet at
8498, 8542, and 8662 \AA.  Both sets of lines increase in strength (in
accord with Figure~\ref{fig:ew}), but the increase in [Ca~{\sc ii}]
emission flux is more dramatic because these lines were very weak on
day 34.  This increase is similar to the increase in [Ca~{\sc ii}]
emission strength seen in light echoes of $\eta$ Car as it faded
\citep{prieto14}. The velocity profiles of the lines also change
markedly between the two epochs.  The [Ca~{\sc ii}] lines develop a
boxy or even double-peaked profile, with a FWHM of 821($\pm$5) km
s$^{-1}$, and with the line wings dropping steeply to the continuum
level at $-$530 ($\pm$15) and +530 ($\pm$15) km s$^{-1}$.  Thus, the
emission is narrower than the broad asymmetric emission profile of
H$\alpha$ at late epochs (Figure~\ref{fig:halpha}). The Ca~{\sc ii} IR
triplet lines show the same boxy emission profile with almost exactly
the same width (see the inset in Figure~\ref{fig:ca2}), but the tops
of the profiles are skewed to the red, presumably because of some
blueshifted self-absorption that diminishes the blue peak.  The
Ca~{\sc ii} IR triplet lines also have blueshifted P Cygni absorption
components, with the velocity of the absorption trough minimum
increasing dramatically from $-$350 ($\pm$30) km s$^{-1}$ on day 34 to
$-$800 ($\pm$20) km s$^{-1}$ on day 1836.  The velocity of the Ca~{\sc
  ii} P Cygni absorption is therefore similar to the speed of the blue
bump in the H$\alpha$ emission profile at late times.  The Ca$+$
emission components do not trace this faster material, and therefore
originate from a different zone in the outflow; this may be relevant
for interpretations of the [Ca~{\sc ii}] emission in other SN
impostors, like the SN~2008S-like objects mentioned in the
Introduction, as well as the [Ca~{\sc ii}] emission seen in light
echoes of $\eta$ Carinae, which have been compared to UGC~2773-OT
\citep{rest12,prieto14}.  The geometry of the outflow is discussed
below in Section 4.2.

\section{DISCUSSION}

\begin{figure*}
\includegraphics[width=6.1in]{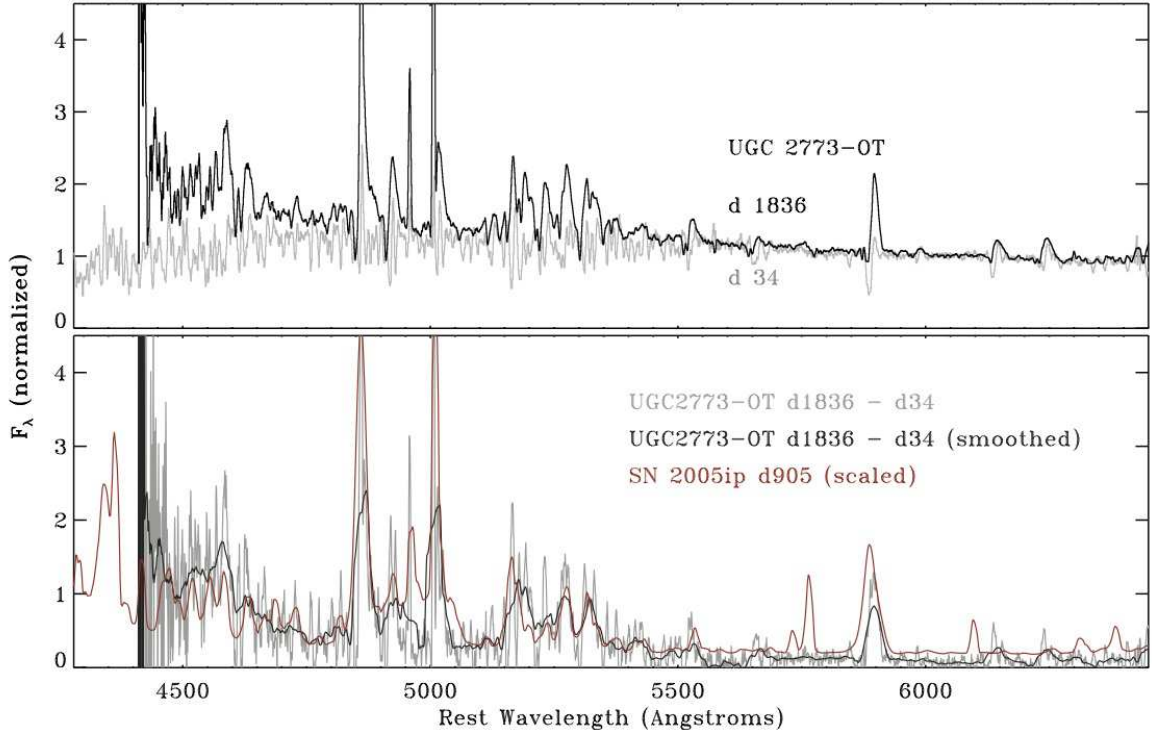}
\caption{A comparison of the early and late-time spectra for
  UGC~2773-OT, highlighting the excess blue emission.  The top panel
  shows the day 1836 spectrum (black) compared to the day 34 spectrum
  from \citet{smith10} in gray. Both spectra are normalised to the red
  continuum level, and in fact the red spectra appear very similar
  except for H$\alpha$ and Ca$^+$.  In the blue there is a clear
  excess of line emission at late times.  The bottom panel isolates
  this excess blue emission by subtracting the normalised day 34
  spectrum from the normalised day 1836 spectrum.  The gray plot is
  the residual, and the black plot is a smoothed version of the
  residual emission.  We compare this to the excess ``blue
  pseudo-continuum'' seen in SN 2005ip (red) on day 905
  \citep{smith09}, which is characteristic of the forest of blue
  emission lines seen in interacting SNe.  This indicates that the
  excess blue line emission in UGC~2773-OT is most likely powered by a
  shock interacting with CSM.}
\label{fig:blue}
\end{figure*}

\subsection{Wind/ejecta vs. Shock Interaction}

While the light curve of UGC~2773-OT shows a very gradual rate of
fading, there are interesting spectroscopic changes over the past 6
years that hold important clues about the physical processes in the
eruption.  Overall, the observed spectrum held an almost constant
continuum temperature at 6500~K while fading about 1.5 mag in visible
light.  Despite the unchanging observed continuum slope, there are
signs of increasing ionization and excitation that seem inconsistent
with this relatively low and constant apparent continuum temperature.
The early time spectra around peak apparent brightness are
qualitatively consistent with expectations for either a cool optically
thick wind, or a photosphere that is formed in the expanding ejecta of
an explosion.  As time proceeds, however, we begin to see clear signs
that a strong shock wave is contributing to the spectrum.  This
spectral evolution therefore gets to the crux of a key current
question regarding LBV eruptions (see \citealt{smith+11}): are they
driven by super-Eddington winds or hydrodynamic explosions?  While we
cannot claim from available evidence that a super-Eddington wind
\citep{owocki04} plays no role in lifting material from the star (in
fact it may help produce the CSM into which the explosion crashes),
there are several lines of evidence that point to or require strong
shock excitation as playing a key role in powering the event,
especially at later times.  These are:

1.  In the past 6 years, H$\alpha$ has increased sharply in EW and
line luminosity in UGC~2773-OT, as seen in the normal evolution of
SNe~IIn powered largely or entirely by CSM interaction.  Also, the
H$\alpha$/H$\beta$ ratio increases as the H$\alpha$ luminosity climbs.
This increasing flux ratio favours collisional excitation in a shock
rather than photoionization and recombination as the power source for
the strong H$\alpha$ emission.  Again, late-time spectra of SNe~IIn
exhibit very strong H$\alpha$, often with little else detected in
their visible-wavelength spectra.

2.  The EW and line flux of He~{\sc i} $\lambda$5876 strengthens after
day 1000, even though the apparent continuum temperature changes
little.  A 6500~K photosphere cannot photoionize a substantial
fraction of the He to produce He~{\sc i} recombination emission.  This
requires an additional source of high excitation.  Interestingly, the
Ca~{\sc ii} IR triplet weakens as the He~{\sc i} $\lambda$5876 EW
strengthens, while H$\alpha$ remains strong after day 1000
(Fig.~\ref{fig:ew}).  This again suggests an increase in ionization
from Ca$^+$ to Ca$^{++}$. A significant change in ionization is not
expected for photoexcitation alone if the apparent continuum
temperature is unchanged.  He~{\sc i} $\lambda$5876 emission generally
would require temperatures above 20,000~K.  The excess ionization is
likely due to X-rays and far-UV radiation from a shock.

3.  Characteristic outflow speeds are increasing with time, as seen in
the increased width of the H$\alpha$ emission profile as well as the
increasing speed of the blueshifted P Cygni absorption trough of
H$\beta$ (from about 300 to 800 km s$^{-1}$) and Ca~{\sc ii}.  If the
outflow speed is increasing, one expects that fast ejecta must catch
up with slow material ejected previously, giving rise to a shock.

4.  The blue portion of the visible spectrum changed from a forest of
narrow absorption lines at early times indicative of the strong
line-blanketing seen in dense winds and cooling ejecta (day 34) to an
excess of blue emission lines at later epochs.  Figure~\ref{fig:blue}
directly compares the day 34 spectrum with strong blue line blanketing
to the day 1836 spectrum with excess blue emission, with both spectra
normalised to the red continuum level.  While their red spectra are
very similar (except for the strength of H$\alpha$ and a few other
individual lines), the spectra at these two epochs are very different
in the blue.  To isolate the {\it excess} blue emission, the bottom
panel in Figure~\ref{fig:blue} shows a spectrum of the residual after
subtracting the day 34 spectrum from the later day 1836 spectrum.  The
residual represents the extra emission that appears at late times.
This residual emission is shown in gray, while the black tracing is a
smoothed version of this same residual spectrum.  This residual blue
emission has a spectral morphology that is very reminiscent of the
``blue pseudo continuum'' that is often seen in SNe~Ibn and SNe IIn
that are powered by strong CSM interaction.  Figure~\ref{fig:blue}
also shows the late-time (day 905) spectrum of SN 2005ip (red) from
\citet{smith09}, which was a Type~IIn explosion that was unusually
bright at late times because of strong and persistent ongoing CSM
interaction. With the exception of a few narrow coronal lines that are
especially strong in SN~2005ip \citep{smith09}, Figure~\ref{fig:blue}
shows that a late-time SN~IIn actually provides quite a good match for
the blue excess emission in UGC~2773-OT.

The presence of fast ejecta and a growing contribution to the observed
spectra from shock emission is interesting in the context of making
comparisons between UGC~2773-OT and the Great Eruption of
$\eta$~Carinae.  There are several lines of evidence that point to a
strong shock as being important for powering $\eta$ Car's Great
Eruption: (1) The very fast ejecta outside the Homunculus, moving at
speeds of 3000-5000 km s$^{-1}$ \citep{smith08}, which are hard to
achieve in a wind model that can also form the slower Homunculus
\citep{owocki04}, (2) the large ratio of kinetic energy to total
radiated energy that exceeds unity \citep{smith03}, (3) the very thin
walls of the Homunculus nebula that seem to have been compressed in a
radiative shock \citep{smith06}, and (4) spectra of light echoes of
$\eta$ Car that seem inconsistent with expectations for an opaque wind
model \citep{rest12,prieto14}.  These indicators of explosive mass
loss seem, at first glance, to be at odds with the decade-long
duration of the bright plateau in $\eta$ Car, which is of course much
longer than the short duration of hydrodynamic mass ejection.  Using a
simple 1-D model of CSM interaction, however, \citet{smith13} showed
that the long plateau of $\eta$ Car's 19th century event could be
powered by CSM interaction by adopting physical parameters that are
consistent with the mass and kinetic energy now observed in the
Homunculus.  The fact that observed spectra of the UGC2773-OT
transient show evidence for an increasing contribution from
shock-powered emission reinforces the conclusion that shocks play an
important role in powering the long-duration high-luminosity plateaus
of these two events, and also bolsters the comparison between
UGC2773-OT and $\eta$~Car (note that spectra of light echoes
corresponding to $\eta$ Car's long 1850s plateau are still being
analyzed; Smith et al., in prep.).

Some observed evidence for strong shock excitation has been reported
for other SN impostors as well, usually showing a characteristically
``hot'' spectrum (see \citealt{smith+11}).  Some of these are the
pre-SN eruptions of SN~2009ip \citep{smith10,foley11}, SN~2000ch
\citep{smith+11,pastorello10,wagner+04}, and SNHunt248
\citep{mauerhan15}.  All of these, however, have exhibited brief
luminosity spikes (weeks to months), not sustained decade-long events
like $\eta$~Carinae and UGC~2773-OT.

\subsection{Asymmetry in the Ejecta}

One of the most remarkable aspects of the spectral evolution of
UGC~2773-OT in its first 2000 days is its H$\alpha$ line
(Figure~\ref{fig:halpha}).  In particular, we note the change in
profile shape from a normal P~Cyg profile that one might attribute to
a moderately slow wind, to a multi-peaked and asymmetric pure emission
profile, accompanied by a huge increase in strength of H$\alpha$
emission.

As noted above, various spectral clues suggest a significant
contribution to the optical emission from shock excited gas in CSM
interaction.  This is especially true for H$\alpha$, where shock
excitation likely dominates the broad ($\pm$1000 km s$^{-1}$)
H$\alpha$ luminosity.  While the earliest epoch showed a similar line
profile in H$\alpha$ and H$\beta$, the late-time H$\beta$ does not
show a broad multi-peaked line profile like H$\alpha$, and it does not
exhibit a comparable increase in strength.  In fact, no other emission
line in the spectrum shows a profile similar to that of H$\alpha$.
This argues that the strong H$\alpha$ emission is caused by shock
excitation in a similar vein to the strong H$\alpha$ emission that
often dominates the late-time spectra of SNe~IIn (e.g.,
\citealt{smith+14}).  In a steady wind, one expects Balmer transitions
and other lines to exhibit similar profile shapes.

\begin{figure}
\includegraphics[width=2.9in]{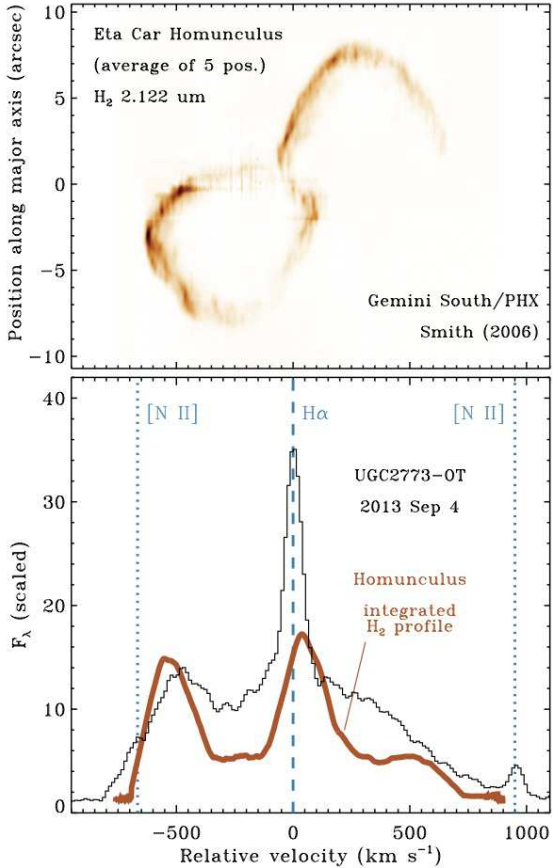}
\caption{Comparison between the asymmetric H$\alpha$ line profile
  observed in UGC~2773-OT to the H$_2$ emission from the Homunculus
  nebula around $\eta$ Carinae.  The top panel shows the 2D long-slit
  spectrum of H$_2$ S(1-0) 2.122 $\mu$m emission from the Homunculus.
  This is an average of 5 adjacent slit positions with the slit
  aligned along the major (polar) axis of the bipolar nebula,
  excluding a region around the bright central star.  These H$_2$
  spectra were obtained with the Phoenix spectrograph on Gemini South,
  and the individual positions were presented by \citet{smith06}.  The
  bottom panel compares the line profile of this H$_2$ emission
  integrated along the slit (meant to mimic the integrated H$_2$ line
  profile observed for the whole Homunculus nebula).  This is the
  thick orange curve, which is compared to the H$\alpha$ profile
  observed in UGC~2773-OT (black) on 2013 Sep.\ 4.}
\label{fig:halphaEta}
\end{figure}

If we assume that late-time H$\alpha$ emission traces the postshock
gas in CSM interaction, then the unusual profile of H$\alpha$ holds
important clues about the geometry of the ejecta and CSM.  The line
became much stronger with time, but the profile shape also changed
dramatically.  A blue bump of excess emission appeared in our second
epoch of spectra on day 478/479 (see Figure~\ref{fig:halpha}).  The
blue bump was located at roughly $-$360 km s$^{-1}$, which is roughly
the same speed as the P Cygni absorption trough in the first epoch on
day 34 \citep{smith10}.  (This blue bump is not at the correct
wavelength to be [N~{\sc ii}] $\lambda$6548; see
Figure~\ref{fig:halpha}.)  The blue bump then grew stronger with time
and migrated to faster blueshifted speeds.  At later times (days
1000-2000) the blue bump is seen at roughly $-$600 to $-$800 km
s$^{-1}$, while a corresponding red bump develops as well (although
less pronounced).  At later times the H$\alpha$ profile exhibits very
steep blue and red wings, unlike the profile wings from electron
scattering in a wind.

A qualitatively similar asymmetric blue bump in H$\alpha$ has been
seen at different speeds in several SNe~IIn
\citep{smith12,smith12b,smith15,fransson14}, and is usually attributed
bipolar or perhaps disk-like geometry in the shock interaction.  A
corresponding red bump is usually assumed to be weaker or absent due
to extinction by dust or occultation by the opaque SN photosphere.
Radiative transfer simulations of SNe~IIn with bipolar CSM support
these expectations of having an asymmetric blue emission line arising
from a bipolar geometry \citep{dessart15}.  The increasing speed with
time seen in UGC~2773-OT is likely due to the dominant emission
migrating from the slower preshock CSM to the accelerated post shock
gas.

This sign of asymmetry in the line profiles during the eruption has
important implications for understanding the origin of bipolar nebulae
around LBVs and other massive stars.  Since the blue bump is seen from
emission lines that arise in CSM interaction during the transient
event itself, this requires that the shaping of the bipolar nebula
occurs during the event or before (imprinted in the inner CSM), and is
not due to hydrodynamic shaping over a long time after the event.  So,
for example, the bipolar nebula seen centuries or thousands of years
later does not arise from an asymmetric posteruption wind, nor from
interaction between a wind and an extended disk-like CSM at large
radii.  The bipolar geometry must be inherent to the physics of the
explosion itself, or to the CSM ejected immediately before that is
overtaken by the ejecta during the observed event.

In this paper we have compared UGC~2773-OT's recent long-lasting
eruption to the historical 19th century eruption of $\eta$ Carinae
\citep{sf11}.  In that spirit, Figure~\ref{fig:halphaEta} makes
another relevant comparison.  This figure shows the asymmetric
H$\alpha$ line profile observed at late times in UGC~2773-OT, and
compares it to the present-day line profile of emission from the
Homunculus nebula around $\eta$ Car.  At the present epoch, $\eta$
Car's ejecta have cooled and the nebula is seen as a dusty reflection
nebula at visible wavelengths, rather than in H$\alpha$ emission.
However, the dense thin shell that contains most of the mass in the
Homunculus emits bright near-IR H$_2$ lines \citep{smith06}.  In the
CSM interaction model proposed by \citet{smith13}, the thin walls of
the Homunculus seen today in H$_2$ correspond to the cold dense shell
of postshock gas that would have emitted bright H$\alpha$ during the
eruption.  Therefore, the comparison in Figure~\ref{fig:halphaEta} is
physically motivated, because the different lines are tracing
essentially the same postshock gas, before or after it cools.  The
H$_2$ emission line profile shown here is an attempt to mimic the
integrated H$_2$ profile of the entire nebula, as if the Homunculus
were an unresolved point source at a large distance.  This is
accomplished by summing the H$_2$ emission along the slit, for the
several different slits used to map the Homunculus with the Phoenix
spectrograph at Gemini South.  These data were originally presented
and discussed by \citet{smith06}.  The result of this exercise is that
the integrated H$_2$ emission from the Homunculus nebula shows a
profile that is very similar to that of the H$\alpha$ emission
observed now in UGC~2773-OT.  A strong blue bump is seen from the
front of the approaching blueshifted polar lobe of the nebula, and a
central peak is seen from slower material at the pinched waist of the
nebula.  The emission peak from the receding polar lobe is suppressed
in this case by dust extinction in the Homunculus
\citep{smith06}. This comparison makes a strong plausibility argument
that the H$\alpha$ profile seen in UGC~2773-OT could arise in a
bipolar shell created by CSM interaction, as in $\eta$ Car, and it
underscores the potential similarity of the two objects.

At late times, lines of [Ca~{\sc ii}] and Ca~{\sc ii} also exhibit
unusual and asymmetric profile shapes (Figure~\ref{fig:ca2}), and the
emission components trace different velocities than H$\alpha$.  The
forbidden lines of the [Ca~{\sc ii}] $\lambda\lambda$7291,7324 are not
contaminated by absorption.  Their profiles are quite boxy, with
double peaked horns at the relatively flat top of the profile, and
their steep line wings extend only to about half the velocity seen in
H$\alpha$. Such differences are not expected in a wind where radiation
scatters out from a central source, and where forbidden lines would
presumably trace the outermost material at the final coasting velocity
(faster than in the inner wind acceleration region).  Instead, the
stark differences in line profile shape and width between H$\alpha$
and [Ca~{\sc ii}] may arise because they trace different latitudinal
zones in the shock front.  A bipolar shock front will have
significantly different speeds -- and hence different
excitation/ionization levels -- at the slow equator and fast poles.
H$\alpha$ may trace the stronger and faster shock at the polar regions
of the bipolar shock, whereas [Ca~{\sc ii}] emission may be restricted
to lower speed and lower excitation material near the equator or mid
latitudes.  The fact that the IR triplet Ca~{\sc ii} lines show these
faster speeds in absortption that are similar to H$\alpha$ emission
supports this conjecture.

\subsection{A Key Difference}

Although we have discussed above how UGC~2773-OT bears many observable
similarities to the historical 19th century eruption of $\eta$
Carinae, there is one key difference that needs to be mentioned.
This difference stems from the fact that UGC~2773-OT shows a smooth
light curve, whereas $\eta$ Car shows a series of repeating brief
luminosity spikes in its historical data \citep{sf11}, especially
before 1845, as noted in the Introduction.

These brief luminosity spikes seem to coincide with times of
periastron passage in the very eccentric binary system if we extend
the present day 5.5 yr orbit back in time, with a small adjustment to
the period for the mass lost in the event \citep{sf11,damineli}.  The
periastron events have been discussed in terms of grazing stellar
collisions at periastron due to an inflated primary star's envelope
\citep{smith11}, and periastron accretion events onto a companion that
power the event and blow jets that shape the bipolar Homunculus
\citep{soker01,ks09}.  If these brief encounters are missing in
UGC~2773-OT but it produces a long-lived eruption and a bipolar nebula
anyway, we may be receiving a clue that the periastron collisions
could be a secondary effect rather than the driving mechanism of these
eruptions.  This last point is speculative, however, and additional
examples of similarly persistent LBV giant eruptions with well-sampled
photometry would be valuable in this regard.

\section*{Acknowledgements}

\scriptsize 

We thank an anonymous referee for a careful reading of the manuscript.
We thank the staffs at the MMT, LBT, Bok, Lick, and Keck Observatories
for their assistance with the observations.  We thank Charles
Kilpatrick for assistance with some of the Kuiper observing, and Betsy
Green for assistance with Bok observations.  Observations using
Steward Observatory facilities were obtained as part of the large
observing program AZTEC: Arizona Transient Exploration and
Characterization.  Some observations reported here were obtained at
the MMT Observatory, a joint facility of the University of Arizona and
the Smithsonian Institution.  This research was also based in part on
observations made with the LBT.  The LBT is an international
collaboration among institutions in the United States, Italy and
Germany. The LBT Corporation partners are the University of Arizona on
behalf of the Arizona university system; the Istituto Nazionale di
Astrofisica, Italy; the LBT Beteiligungsgesellschaft, Germany,
representing the Max-Planck Society, the Astrophysical Institute
Potsdam and Heidelberg University; the Ohio State University and the
Research Corporation, on behalf of the University of Notre Dame,
University of Minnesota and University of Virginia.  Some of the data
presented herein were obtained at the W.M.\ Keck Observatory, which is
operated as a scientific partnership among the California Institute of
Technology, the University of California, and NASA; the observatory
was made possible by the generous financial support of the W.M.\ Keck
Foundation. The authors wish to recognise and acknowledge the very
significant cultural role and reverence that the summit of Mauna Kea
has always had within the indigenous Hawaiian community.  We are most
fortunate to have the opportunity to conduct observations from this
mountain.  Based in part on observations obtained at the Gemini
Observatory, which is operated by AURA, under a cooperative agreement
with the NSF on behalf of the Gemini partnership: the National Science
Foundation (US), the Particle Physics and Astronomy Research Council
(UK), the National Research Council (Canada), CONICYT (Chile), the
Australian Research Council (Australia), CNPq (Brazil), and CONICET
(Argentina). Research at Lick Observatory is partially supported by a
generous gift from Google.

N.S.'s research on Eta Carinae and related eruptions receives support
from NSF grants AST-1312221 and AST-1515559, as well as from NASA
grant AR-12618 from the Space Telescope Science Institute, which is
operated by Associated Universities for Research in Astronomy, Inc.,
under NASA contract NAS 5-26555.  The supernova research of A.V.F.'s
group at U.C. Berkeley is supported by Gary \& Cynthia Bengier, the
Richard \& Rhoda Goldman Fund, the Christopher R. Redlich Fund, the
TABASGO Foundation, and NSF grant AST-1211916. KAIT and its ongoing
operation were made possible by donations from Sun Microsystems, Inc.,
the Hewlett-Packard Company, AutoScope Corporation, Lick Observatory,
the NSF, the University of California, the Sylvia \& Jim Katzman
Foundation, and the TABASGO Foundation.

\end{document}